\documentstyle[12pt]{article}
\begin{document}
\begin{flushright} {OITS 694}\\
July 2000
\end{flushright}
\vspace*{1cm}

\begin{center} {\Large {\bf Observable measures of critical behavior in
high-energy nuclear collisions}}
\vskip .75cm
 {\bf  Rudolph C. Hwa}
\vskip.5cm

{Institute of Theoretical Science and Department of Physics\\
University of Oregon,
Eugene, OR 97403-5203, USA}

\end{center}

\begin{abstract}
Critical behaviors of quark-hadron phase transition in
high-energy heavy-ion collisions are investigated with the aim
of identifying hadronic observables. The surface of the plasma
cylinder is mapped onto a 2D lattice. The Ising model is used to
simulate configurations corresponding to cross-over transitions
in accordance to the findings of QCD lattice gauge theory.
Hadrons are formed in clusters of all sizes. Various measures are
examined to quantify the fluctuations of the cluster sizes and of
the voids among the clusters. The canonical power-law behaviors
near the critical temperature are found for appropriately chosen
measures. Since the temperature is not directly observable,
attention is given to the problem of finding observable
measures. It is demonstrated that for the measures considered the
dependence on the final-state randomization is weak. Thus the
critical behavior of the measures proposed is likely to survive
the scattering effect of the hadron gas in the final state.
\end{abstract}

\section{Introduction}

The subject of this talk can be introduced by commenting on the
four conceptual items mentioned in the title, but in reverse
order.  First, high-energy nuclear collisions.  The goal is to
create quark-gluon plasma.  So far the conventional signatures are
aimed at detecting the primordial quark matter. Suppose that is
found.  Then what?  If we can learn from the condensed-matter
physics, then we know that critical phenomena form a vibrant
area of research.  That leads to the second item in the title: 
critical behavior.  Our concern will be the quark-to-hadron
phase transition (PT).  The hadron-to-quark PT is, in my view,
too turbulent in heavy-ion collisions to lend itself to
systematic studies, despite some recent attempt to use
percolation as a model to characterize it.  The third item in
the title is ``measures.''  For a thermal system the heat
capacity can be a measure that exhibits the critical behavior
$C \propto \left|T-T_c \right|^{-\alpha}$, while for a magnetic
system the magnetic susceptibility behaves as $\chi\propto
\left|T-T_c \right|^{-\gamma}$.  What is the measure for
quark-hadron PT?  Is there a measure, call it $\mu$, which
behaves as $\mu \sim \left|T-T_c \right|^{-\kappa}$?  If so, how
do we find it?  That brings us to the last item:  observable. 
Temperature is not directly observable.  (The transverse
momentum $p_T$ is an unreliable, gross indicator of the average
phenomenon.)  Moreover, the system is not static and the emitted
hadrons may undergo final-state interactions that can smear the
signature that is to be observed.   The aim of this talk is to
show how all these issues are addressed and to indicate where
the answers are to be found.

The measure that I believe will convey the information about
quark-hadron PT is about fluctuations, not only bin-to-bin
fluctuations, but also event-to-event fluctuations.  A way to
illustrate this belief is to consider the annual precipitation
(AP) of two places with widely different climatic
characteristics.  The AP of certain regions in Oregon and
Florida are very similar, about 40$^{\prime\prime}$ per year, but
their rainfall patterns are nearly opposite:  drizzly in Oregon,
stormy in Florida.  The problem is that in determining the PA the
daily rainfall is integrated over the entire year, and then
averaged over all years.  That is exactly how the average
multiplicity of heavy-ion collisions is determined:  integrated
over the whole hadronization time of an event (which can take
over 30 fm/c), and then averaged over all events.  Clearly, to
learn about the nature of rainfall or the properties of
hadronization, it is necessary to study spatial and temporal
fluctuations.

\section{The Problem}

If quark-gluon plasma is indeed formed in a heavy-ion collision,
it is reasonable to assume that the plasma occupies a
cylindrical volume, expanding rapidly in the longitudinal
direction, and more slowly radially.  The medium most likely has
a temperature profile that is high at the center of the
transverse plane and decreases with increasing radius due to the
radial expansion and cooling.  When $T$ reaches $T_c$ at the
surface, the quark matter undergoes a $PT$ into hadrons.  Thus
the characteristics of critical behavior are to be found on the
cylindrical surface at any given time of the evolutionary
history of the system.  That is like raining at a particular
instant of observation.  If we collect all the hadrons emitted
throughout the long hadronization processes, the features of
critical behavior are likely to overlap and get averaged out.  It
is therefore important to select a narrow $\Delta t$ cut in the
hadronization time.  If the emission time and the transverse
momentum $p_T$ of the hadrons are correlated in a one-to-one
relationship, then a narrow $\Delta p_T$ cut would accomplish
what we need.  Unfortunately, that relationship is not
one-to-one.  On the contrary, at each emission time the emitted
particles have a very wide $p_T$ distribution.  That wide
distribution turns out to provide us with a way to proceed. 
Suppose we make a narrow $\Delta p_T$ cut and study the hadronic
patterns in the $\eta$-$\phi$ 2D space orthogonal to
$\vec{p}_T$.  Then at any emission time only a small fraction of
the emitted hadrons would enter the narrow $\Delta p_T$ window. 
This random selection process at each time frame thus
decorrelates the hadrons collected over the entire hadronization
process.  The measured hadronic patterns in the $\eta$-$\phi$
space is then the superposition of small parts of many
configurations (which we call configuration mixing), each of
which we shall stimulate before mixing.  The object of our
analysis is to find the scaling property of each configuration
that exhibits the critical behavior, and then to see whether it
survives the mixing process.

It is well known that a system at the critical point forms
clusters of all sizes that exhibit Kadanoff scaling.  The
details of the dynamical system are not important, since the
critical behavior depends mainly on the dimension and symmetry
of the system.  On the basis of that universality we shall use
the Ising model as a simple device to simulate the configuration
of hadronic clusters formed on the surface of the plasma
cylinder at each time frame of hadronization.  There are
reasons to believe that the QCD dynamics for the chiral PT
suggests a cross-over near the second-order PT that belongs to
the same universality class as the Ising system \cite{kr}.  We
shall define hadron formation on the Ising lattice in such a way
as to be able to simulate a cross-over.  From the simulated
configurations we can see the hadronic clusters randomly
scattered over a background of voids.  In our first-attempt
analysis of the patterns formed, we shall ignore the
complications that arise from the $p_T$ distribution and
consider directly the scaling properties of the clusters and
voids in uncorrelated configurations that we shall simulate.

\section{Cluster Analysis}

The use of the Ising model on a 2D lattice to simulate
second-order PT is standard \cite{bdf,zc}.  To have a cross-over
in quark-hadron PT without the use of an external field in the
Ising Hamiltonian is not standard \cite{hwz}.  The method uses a
cell on the lattice, of dimension $\epsilon \times \epsilon$
where $\epsilon$ may be taken to be 4, to define hadron density
$\rho_c$ at the {\it c}th cell 
\begin{equation}
\rho_c = \lambda  \, \phi^2_c \, \theta\left(\phi_c\right)
\label{1}
\end{equation}
where
\begin{equation}
\phi_c = \sum_{j \in c} \mbox{sgn} \left(m_L \right)
 \sigma_j \quad , \qquad \qquad m_L = \sum_{j \in L^2} \sigma_j
\quad .
\label{2}
\end{equation}
The first sum in (\ref{2}) is over all sites in a cell having
site-spins $\sigma_j$; the second sum is over all lattice sites,
giving the total magnetization $m_L$.  We use the direction of
$m_L$ of each configuration to serve as the direction of an
external field and define the order parameter $\phi_c$ as the
cell spin at $c$ relative to that direction.  $\lambda$ in
(\ref{1}) is a parameter relating hadron density to Ising
spins; all measurable quantities should be insensitive to
$\lambda$.  Because of the $\theta$-function in (\ref{1}),
$\rho_c$ is either positive where hadrons are emitted, or zero
where a void exists.

There are two quantities we have found that can effectively
serve as measures of fluctuations of hadronic clusters from bin
to bin and from configuration to configuration.  One is $J(M,T)$
defined 

\begin{equation}
J(M,T) = \left<{\rho_k  \over  \left<\rho\right>}  \ell n
{\rho_k 
\over \left<\rho\right>  }\right>
\label{3}
\end{equation}
and the other is $K(M,T)$ whose definition is omitted here, but
can be found in Ref. \cite{hwu}.  In (\ref{3}) $\left<
\cdots \right>$ refers to averaging over all bins [$M$ of
them, $M = (L/\delta)^2$] and over all configurations.  We
discuss only the properties of $J(M,T)$ here; those of $K(M,T)$
are similar.
 
What we have found about $J(M,T)$ is that it is factorizable
\cite{hwu}, i.e.,   
\begin{equation}
J(M,T) = \alpha (T) J_c (M) \quad ,
\label{4}
\end{equation}
where $J_c (M) = J(M, T_c)$.  Thus, by definition,
$\alpha \left(T_c\right) = 1$.  Yet, when $T$ is away from $T_c$,
but not too far in $T < T_c, \alpha (T)$ satisfies the power-law
behavior
\begin{equation}
\alpha (T) \propto \left(T_c - T \right)^{-\zeta}, \qquad \qquad 
\zeta = 1.88
\quad .
\label{5}
\end{equation}
This may be regarded as the critical behavior that we have been
searching for.  However, $T$ is not directly measurable.  We can
use $\bar{\rho}$ instead, where $\bar{\rho}$ is the cell density
averaged over all cells in one configuration.  Then we find
\begin{equation}
J(M, \bar{\rho}) = \alpha \left(\bar\rho \right) J_0
(M)
\label{6}
\end{equation}
with 
\begin{equation}
\alpha \left(\bar\rho \right) \propto \left(\bar{\rho}-
\bar{\rho}_0
\right)^{-\bar\zeta}, \qquad \bar\zeta = 0.97 \quad ,
\label{7}
\end{equation}
where $\bar{\rho}_0$ is the average density where
$J(M, \bar{\rho}_0) = J_0(M)$ is the highest.  The
behavior in (\ref{7}) is measurable and can be used to signal
the fluctuation property of a quark-hadron PT.

\section{Void Analysis}

In addition to finding the critical behavior of the hadronic
clusters, we can also analyze the scaling properties of the
voids where no hadrons appear \cite{hwz}.  First, it is
necessary to define precisely what a void is, and how to avoid a
single hadron from changing its quantification.  To that end we
define a bin to be empty if the average density $\bar\rho_k$ of
the {\it k}th bin is less than $\rho_0$, where $\rho_0$ is a
parameter that can be varied by the analyst of the data.  We do
not set $\rho_0$ to zero in order to eliminate the small
fluctuations due to irregular occurrences of hadrons in or
at the edges of a bin.  A void is then the area of
contiguous empty bins, and its area is therefore
\begin{equation}
V_n = \sum_k \theta \left(\rho_0 - \bar\rho_k\right) \quad ,
\label{8}
\end{equation}
where the sum is over all connected empty-bins of the {\it
n}th void.  $V_n$ is in the unit of number of bins.  A
quantity less dependent on the size of the bins is the
fraction $x_n = V_n/M$.  For every configuration, there
exists a set of void fractions $\left\{x_1, x_2, x_3, \dots x_m 
\right\}$ where $m$ is the total number of voids.  The
comparison of void patterns of different configurations would
be difficult if they are characterized by those sets.  To
ease that comparison, we define the $G$ moments
\begin{equation}
G_q = {\left<x^q \right>  \over  \left<x \right>^q} \quad ,
\label{9}
\end{equation}   
where $\left<x^q \right>  = \left(\sum_{n=1}^m \,
x_n^q\right)/m$, an average performed for each configuration. 
Thus $G_q$ is a number (for each $q$) that characterizes the
void pattern of a configuration.  At the critical point this
$G_q$ fluctuates widely from configuration to configuration. 
It depends on the bin size.

At $T$ in the vicinity of $T_c$ we find that $\left<
G_q\right>$, averaged over all configurations, satisfies scaling
law for a variety of $\rho_0$ values, i.e., 
\begin{equation}
\left< G_q\right> \propto M^{\gamma_q} \quad .
\label{10}
\end{equation}
That means voids of all sizes occur.  Furthermore, the scaling
exponents depend on $q$ linearly
\begin{equation}
\gamma_q = c_0 + cq
\label{11}
\end{equation}
with $c = 0.8$ at $T_c$ and $\rho_0 = 20$.  This value of
$\rho_0$ is less than 8\% of the maximum $\bar\rho$.  The value
of $c$ depends on $T$ and $\rho_0$ mildly, and can be used to
quantify the nature of void patterns.

Since $G_q$ fluctuates very widely, i.e., the distribution
$P(G_q)$ is very broad, the moments of $G_q$ can reveal that
fluctuation.  We have studied the derivative of the first
moment, which is
\begin{equation}
S_q = \left< G_q \ell n G_q \right>
\label{12}
\end{equation}
and found that it also satisfies a scaling law
\begin{equation}
S_q \propto M ^{\sigma_q}
\label{13}
\end{equation}
whose exponent depends on $q$ linearly
\begin{equation}
\sigma _q = s_0 + sq
\label{14}
\end{equation}
with $s = 0.76$ at $T = T_c$ and $\rho_0 = 20$.  The value of
$s$ also depends on $T$ and $\rho_0$ mildly, and therefore can
further be used to quantify the nature of the fluctuation of
void patterns from configuration to configuration.

The dependences of $c$ and $s$ on $\rho_0$ and $T$, though mild,
can be used to diagnose the temperature range of quark-hadron
$PT$.  For $\rho_0 > 20$ the values of $c$ and $s$ determined
in our analysis vary in a range as much as 20\% for $T$ varying in
the range of
$\pm 2\%$ around $T_c$, but they converge to the values quoted as
$\rho_0
\rightarrow 20$.  If $c$ and $s$ are not unique for a range of
$T$ around $T_c$, it means that $\gamma_q$ and $\sigma_q$ are
not fixed (for fixed $q$) if PT takes place over a range of $T$,
not exactly at $T_c$.  Then there would be no scaling behavior. 
Thus, if the experimental data reveal the loss of scaling as
$\rho_0$ is increased, it implies that the $PT$ takes place over
a range of $T$.  If, however, scaling persists, then $PT$ occurs
only at one $T$, presumably $T_c$.

\section{Complications in Hadronization}

There are two aspects about hadronization in heavy-ion
collisions that one should be concerned about for the type of
measures we are proposing.  One is the effect due to final-state
interaction as the hadrons traverse the hadron gas.  The other
is the effect of configuration mixing.  The former is studied by
requiring the produced hadron in each configuration to take
$\nu$ steps of random walk, since scattering by a hadron gas
implies the random deflections of the initial $\vec{p}_T$ that
result in random wandering in the $\eta$-$\phi$ plane.  We have
studied the effect of such random walks (for $\nu$ up to 6) on
$J(M,T)$ and found that the factorizability (\ref{4}) persists
and that $\alpha (T)$ retains the essential character of
(\ref{5}) \cite{hwu}.

The other complication of configuration mixing due to the
non-correspondence between $\Delta p_T$ and $\Delta t$ is more
difficult to treat.  We have considered only the simple case
of no correlation between successive time frames.  We mix four
independently simulated configurations by taking one quadrant
from each and then apply our void analysis to the mixed
configurations \cite{hwz}.  We find that the effects on $\gamma
_ q$ and $\sigma _q$ are essentially negligible.  This problem
will be followed by a more realistic treatment of the
configuration mixing where the wide $p_T$ distribution of the
emitted particles at any given time will be taken into account
together with the hadron-void correlation in successive time
frames.

\section{Multifragmants}

Finally, let me make some remarks about ``Common Aspects'' that
is a part of the title of this session.  Despite the valiant
attempts made by the organizers of this conference to promote
interaction between the two subfields encompassed here,
cross-fertilization has not been a vibrant phenomenon that has
taken place.  However, it seems that the ideas about the void
analysis discussed here can be translated into the study of
multifragmentation in the intermediate mass range.  Let me
therefore venture a suggestion on an analysis of the fragments
with the hope that it may yield a feature of the experimental
results more challenging to the model builders than what they
have hitherto dealt with.

Suppose that for each event an experiment can measure a certain
number of fragments whose total mass may be less than the total
mass of the nuclei in collision and may fluctuate from event to
event.  Nevertheless, for all the fragments $M_i$ that have
been detected, one can calculate the fraction $x_i = M_i/\sum_i
M_i$.  If it is the charges of the fragments that are measured,
then let $x_i = Z_i/\sum_iZ_i$.  Each event is then
characterized by a set of fractions $\left\{x_1, x_2, x_3, \dots
x_n 
\right\}$.  The idea is to go beyond studying the averages and
examine the nature of the fluctuation of the fragments
detected.  Let us define
\begin{equation}
H_q(n) = {1  \over  n} \sum^n_{i = 1} x^q_i
\label{15}
\end{equation}
where $n$ is the total number of fragments measure in an event. 
Clearly, $H_0 = 1$ and $H_1 = 1/n$.  At higher $q$, $H_q(n)$ are
progressively smaller, but are increasingly more dominated by the
large $x_i$ components.

If $P\left(H_q \right)$ is the distribution of $H_q$ after all
events are sampled, then the average is
\begin{equation}
\left< H_q\right> = {1   \over  {\cal N}} \sum ^{\cal N}_{e = 1}
H^e_q(n) = \int dH_q \, H_q \, P(H_q)
\label{16}
\end{equation}
where $e$ labels an event and ${\cal N}$ is the total number of
events.  $\left< H_q\right>$
 provides some average information about the multifragments, but
to study the fluctuation from event to event one should
calculate the normalized moments of moments \cite{zch}
\begin{equation}
C_{p,q} = \left< H_q^p\right>/\left< H_q\right>^p
\label{17}
\end{equation}
This $C_{p,q}$ may contain too much information.  The derivative
at $p = 1$ may already be sufficient to provide some interesting
insight into the erratic nature of the fluctuation
\begin{equation}
E_q = \left. {d  \over  dp} C_{p,q}\right|_{p=1} = \left<
{H_q  \over \left<H_q \right> } \ell n{H_q  \over  \left<H_q
\right>}\right> \quad ,
\label{18}
\end{equation}
the study of which may be called erraticity \cite{rch}, for
convenience.  The experimental determination of $E_q$ as a
function of $q$ may pose a serious challenge to the theoretical
models, some of which may not contain enough dynamical
fluctuations to generate the erraticity revealed by the data.

\section{Acknowledgments}
The work described here was carried out at different stages in
collaboration with Z.\ Cao, Y.\ F.\ Wu and Q.\ H.\ Zhang.   It
was supported, in part, by the U. S. Department of Energy under
Grant No. DE-FG03-96ER40972.

\end{document}